\documentclass[11pt]{article}
\usepackage{graphicx}

 \advance\textheight by \topskip
\large\normalsize
\newcommand{\be}{\begin{equation}}
\newcommand{\ee}{\end{equation}}
\newcommand{\ba}{\begin{array}}
\newcommand{\ea}{\end{array}}
\newcommand{\pa}{\partial}
\def\simleq{\; \raise0.3ex\hbox{$<$\kern-0.75em \raise-1.1ex\hbox{$\sim$}}\; }
\def\simgeq{\; \raise0.3ex\hbox{$>$\kern-0.75em \raise-1.1ex\hbox{$\sim$}}\; }

\begin{document}

\begin{center}
{\Large \bf Relic Backgrounds of Gravitational Waves from Cosmic Turbulence}\\
\vskip1.cm
{Alexander D.~Dolgov$^1$, Dario Grasso$^2$ and Alberto Nicolis$^2$}\\
\vskip0.5cm
$^1$ {\em INFN, sezione di Ferrara, Via del Paradiso 12, 
44100 Ferrara, Italy\\
and ITEP, Bol.~Cheremushkinskaya, 25, 117259, Moscow, Russia}\\
$^2$ {\em Scuola Normale Superiore, P.zza dei Cavalieri 7, 56126 Pisa, Italy\\
and INFN, sezione di Pisa}

\begin{abstract}
Turbulence may have been produced in the early universe during several kind
of non-equilibrium processes. Periods of cosmic turbulence may have left 
a detectable relic in the form of stochastic backgrounds of
gravitational waves. In this paper we derive general expressions
for the power spectrum of the expected signal. Extending previous
works on the subject, we take into account the effects of a continuous
energy injection power and of magnetic fields. Both effects lead 
to considerable deviations from the Kolmogorov turbulence spectrum. 
We applied our results to determine the spectrum of gravity 
waves which may have been produced by neutrino inhomogeneous diffusion
and by a first order phase transition. We show that in both cases 
the expected signal may be in the sensitivity range of LISA.
\end{abstract}

\end{center}

\section{Introduction}

The high isotropy of the cosmic microwave background radiation 
(CMBR) testifies that the universe was a quiet
child at photon decoupling time.
This does not prevent it, however, to have been a quite
{\em turbulent} baby!
 
Several violent and interesting phenomena 
may have taken place before the universe became matter dominated.
Phase transitions, and re-heating at the end of inflation
are  examples (we will discuss another one below)
of processes which may have taken the universe through a phase 
during which thermal equilibrium and homogeneity can be temporarily lost. 
Turbulence is expected to have developed under such conditions
due to the huge value of the Reynolds number. 
It would be extremely interesting for cosmologists 
to detect any observable relic of such turbulent periods
as this may shed light on the early stages of the universe and
on fundamental physics which is not yet probed in the laboratories.
Unfortunately, due to the finite thickness of the last scattering surface,
density or metric perturbations which may have been produced by
any causal process  before the matter radiation equality time
than can hardly give rise to detectable imprints on the CMBR. 
The detection of cosmological background(s) of gravitational 
waves (GWs) may offer the only possibility to probe 
turbulence in the remote past.  Although any direct detection of
GWs is still missing,  this subject is not any more purely academic
as a new class of ground and space based observatories of GWs
are under construction or in advanced project. 
The amplitude and frequency sensitivity of some of these instruments
are in the proper range to probe many  interesting astrophysical and 
cosmological processes \cite{maggiore}. The most interesting project from our point
of view is LISA (Laser Interferometer Space Antenna)
which is scheduled to be launched around the end of this
decade and will possibly achieve a sensitivity in $h_0^2 \Omega_{\rm gw}$ of $10^{-12}$ 
at milliHertz frequencies \cite{LISA}.

So far the interest on GWs produced by cosmological turbulence 
has been mainly stimulated by the possibility that turbulence
could have been generated at the end of the electroweak phase transition (EWPT)
\cite{KKT,KMK,alberto}. The idea is that, if the EWPT
is first order, turbulence should have been produced when the
expanding walls of bubbles containing the broken symmetry phase,  
or the shock waves preceding them, collide at the end of the
transition. As discussed in \cite{KMK,alberto} the 
stochastic background of GWs produced during this process may be
detectable by LISA. 
   
Recently two of us \cite{dario} noticed that another, possibly
detectable, background of GWs  may have been produced before neutrino
decoupling, that is much later  than the EWPT.  
This mechanism requires that the net lepton number density
($N_a({\bf x}) \equiv n_{\nu_a}({\bf x}) - n_{\bar \nu_a}({\bf x})$,\ \ $a =
e,\mu,\tau$) of one, or more, neutrino species was not uniform
and changed in space over some characterist scale which was smaller  
than the Hubble horizon at the neutrino decoupling time. 
A net flux of neutrinos should then be produced along the lepton number
gradients when the neutrino mean free path became comparable to
the size of the fluctuations in the neutrino number. It was shown
that, depending on the amplitude of the fluctuations, such a flux of neutrinos
may be able to stir chaotically the cosmic plasma producing
magnetic fields and GWs.  Fluctuations in the neutrino number, as
those which are required to power all this rich set of effects, 
can be a by-product of leptogenesis at the GUT scale  
(this is possible for example in the Affleck-Dine \cite{DolKir}
scenario  of leptogenesis) or be a consequence of active-sterile
resonant neutrino conversion before neutrino decoupling \cite{DiBari}.    

The aim of this paper is to estimate the characterists amplitude 
and spectral distribution of GWs produced  by the mechanism discussed
in \cite{dario} and during a first order phase transition.
Our treatment follows that of Kosowsky, Mack and Kahniashvili
\cite{KMK} (KMK) and extend it by introducing some 
physically relevant generalizations. 

In Sec.~\ref{turbo} we will show how the Kolmogorov spectrum of 
turbulence (which is that adopted by KMK) is modified when energy
injection into the primordial plasma takes place over a continuos range of scales. 
This is motivated by the observation that  neutrino number
fluctuations, which might power turbulence, generally will not have a unique
size but have a more complex spectral distribution. 
The same consideration can be naturally applied to other kind of stochastic sources.
Since strong magnetic fields are expected to be produced by
the mechanism presented in Ref.~\cite{dario} and during first order 
phase transitions \cite{Baym,report}, in Sec.~\ref{MHD} we will
also investigate possible Magneto-Hydro-Dynamical (MHD) effects on the
turbulent spectrum. We will see that these kind of effects may give rise
to quite drastic consequences on the expected GW signal.
In Sec.~\ref{GWs} we apply the results of the
previous two sections to determine the general expression 
of the GW power spectrum produced by cosmological turbulence. 
In Sec.~\ref{neutrino} we estimate the 
signal to be expected at LISA  because of GWs generated by neutrino
inhomogenoeus diffusion. In Sec.~\ref{EWPT} we will determine 
the characteristics of the GW signal produced by a first order phase
transition paying particular attention to the case of the EWPT.
We will show that MHD effects may play a crucial role by enhancing the
expected signal. Finally Sec.~\ref{end} contains our conclusions.

\section{Turbulence spectrum from stirring at any scale}\label{turbo}

Three-dimensional turbulence shows a cascade of energy from larger length scales to smaller ones
({\em direct} cascade): largest eddies form at the scale at which the stirring source acts and
after a few revolutions they break down into smaller eddies; this process goes on until the 
{\em damping scale} is reached (see below).
Call $\varepsilon_k$ the rate at which a (momentum) scale $k$ receives turbulent energy
(normalized to the enthalpy $w \equiv \rho + p$ of the fluid)
\be
\varepsilon_k \equiv \frac{1}{(\rho+p)} \frac{(d \rho_k)_{\rm
in}}{dt} \sim \frac{(d u_k ^2)_{\rm in}}{d t}    \; ,
\ee
where $u_k$ is the typical fluid velocity at the scale $k$, and we
assumed constant energy density $\rho$ and pressure $p$, {\em i.e.}~that
the fluid is incompressible. If the process is (almost)
stationary, $\varepsilon_k$ must equal the rate at which energy is
transferred to smaller scales, {\em i.e.}
\be     
\label{rate}
\varepsilon_k \sim u_k^3 k ~,
\ee
where we assumed (as it is usually done) that the rate at which an eddy breaks down into 
smaller ones is roughly its turnover angular frequency $\omega_k \sim k u_k$.

In general  stirring of the fluid by an external source does not take place at a single 
scale. Rather, turbulence in the early universe is likely to be produced by 
stochastic sources injecting energy over a finite range of wavenumbers. 
Consider any momentum $k$ within this interval. An infinitesimally larger momentum 
$k+dk$, apart from the energy from the turbulent cascade, will receive an additional 
external power $P_{\rm ext} (k) dk$, due to the external stirring at  $k+dk$, that is
\be \label{pot_ext}
\frac{d \varepsilon_k}{d k} = P_{\rm ext} (k)   \; .
\ee
This is different from conventional Kolmogorov type turbulence where energy is supposed to be
injected at a single scale and $\varepsilon_k = \varepsilon$ is momentum
independent.

Following a standard approach (see e.g.~\cite{frost,Biskamp} for a pedagogical introduction to
turbulence) we define the {\em energy spectrum} $E(k)$ by
\be \label{def_spectrum}
\langle {\bf u} ^2 ({\bf x}) \rangle \equiv \int_0 ^\infty E(k) dk  \; ,
\ee
with $\langle \cdots \rangle$ standing for ensemble averages; by combining
eqs.~(\ref{rate}--\ref{def_spectrum}), and noticing that eddies satisfy the
``uncertainty'' relation $\Delta k / k \sim 1$
\footnote{ Properties of Fourier transforms automatically include the uncertainty relation
$\Delta x \Delta k \sim 2 \pi$; moreover, it is clear that the spatial width $\Delta x$ of an eddy
is roughly its characteristic wavelength $\lambda = 2 \pi/k$. This gives $\Delta k / k \sim 1$.
}, we find the spectrum
\be \label{E_di_k}
E(k) \sim \left[ \int_{k_S} ^k P_{\rm ext}(k')dk'\right]^{2/3} k^{-5/3} \; ,
\ee
where $k_S$ is the largest length scale at which we inject energy in turbulent motions.
Note that eq.~(\ref{E_di_k}) reduces to the usual Kolmogorov spectrum
for a delta-like stirring spectrum $P_{\rm ext}$.

The energy cascade stops at the {\em damping scale} $k_D$, characterized by a {\em local}
Reynolds number $\displaystyle {\rm Re}_D \equiv \frac{u_{D}}{k_D \: \eta}$ of order unity, {\em i.e.}
\be
k_D \sim \varepsilon _{\rm tot} ^{1/4} \: \eta ^{-3/4}  \; ,
\ee
where $\varepsilon_{\rm tot} \equiv \int P_{\rm ext}(k)dk$ is the {\em total}
external stirring power.
The energy spectrum $E(k)$ has $k_D$ as ``ultraviolet'' cutoff since
any energy injected at $k > k_D$ is dissipated into heat.

We expect substantial deviations from Kolmogorov spectrum if the integral
appearing in eq.~(\ref{E_di_k}) is substantially increasing with $k$; in such cases we
parametrize it as a power law function of $k$ for $k \gg k_S$,
\be \label{eps_k}
\varepsilon_k = \int _{k_S} ^k P_{\rm ext}(k')dk' = \varepsilon_S \left( \frac{k}{k_S}
\right)^{3\gamma/2}      \; ,
\ee
with $\gamma>0$;
this gives
\footnote{Eqs.~(\ref{E_di_k}) and (\ref{E_di_k_3}) have to be intended as a rough
        estimate of the real physical spectrum. Although they should give a
        good approximation for small values of $\gamma$ (which corresponds to
        some sort of adiabaticity in momentum space) they are probably very crude
        for $\gamma$ of the order of unity. We thank  G.~Falkovich for
        making us this comment.} 
\be     \label{E_di_k_3}
E(k) \sim \varepsilon_S ^{2/3} \: k_S ^{-\gamma}
\: k^{\gamma -5/3}   \; .
\ee
Again, Kolmogorov spectrum would correspond to $\gamma=0$.

It is helpful to rewrite the spectrum and all the other quantities
parametrizing them with the largest-scale characteristic velocity $u_S$ rather than
with the characteristic rate $\varepsilon_S$.
This permits a more intuitive and direct generalization to the relativistic case (see below).
From eqs.~(\ref{rate}) and (\ref{eps_k}) we obtain
\be \label{u_k}
u_k \sim \left( \frac{\varepsilon_k}{k} \right)^{1/3} = u_S
\left( \frac{k}{k_S} \right)^{\gamma/2-1/3}  \; ,
\ee
where $u_S = (\varepsilon_S / k_S)^{1/3}$. We thus have
\begin{eqnarray}
\omega_k ~ & \! \sim \! & \! ~u_S \: k_S ^{1/3 - \gamma/2} \: k^{\gamma/2 + 2/3}  \label{omega_k}\\
\label{E_di_k_4}
E(k) ~ & \! \sim \! & \! ~u_S ^2 \: k_S ^{2/3-\gamma} \: k^{\gamma -5/3}  \\
k_D ^{4/3 - \gamma/2}~ \! & \! \sim \! & \! ~\frac{u_S \: k_S ^{1/3 - \gamma/2}}{\eta}  \label{kD}
\; .
\end{eqnarray}

Notice that for $\gamma>2/3$ the characteristic velocity $u_k$
increases with $k$: this means that even if on large scales the
plasma has non-relativistic turbulent velocities, it will reach
relativistic velocities at a certain $k_R$, if the latter is
smaller that the damping wave number $k_D$. In such a case the
characteristic turbulent velocity for $k>k_R$ will be saturated at
$u_k \sim 1$ \footnote{More precisely, $u_k$ will saturate to the
sound velocity $c_s$ which is $1/\sqrt{3}$ for a relativistic
fluid. The excess of energy injected into the fluid will be
converted into supersonic shock waves. Although the approximation
to treat the fluid as incompressible clearly fails in this case,
it is reasonable to assume that the presence of shock waves will
only make more involved the energy injection process without
affecting turbulence properties. Notice that $u_k$ will be always
smaller than $c_s$.}: clearly this will be the most
interesting case for the production of gravitational waves. In
order to extend to that relativistic case all the formulae we will
derive throughout the paper, it is sufficient to replace $\gamma$
with 2/3 and $u_S$ with 1: this is straightforward to understand
by noticing that with those formal replacements eq.~(\ref{u_k})
produces a costant velocity spectrum with $u_k$ equal to 1. In
particular, by applying that prescription to
Eqs.~(\ref{omega_k}--\ref{kD}) we obtain
\be
u_k \sim 1 \Rightarrow \left\{ \begin{array}{l}
    \omega_k \sim k     \\
    E(k) \sim k^{-1}    \\
    k_D \sim 1/\eta
                   \end{array}  \right.     \; .
\ee

A final remark to this section is in order.  The standard theory 
of turbulence, on which we based our previous considerations, has 
been formulated and tested only for non-relativistic fluid velocities.
A relativistic theory of turbulence is, unfortunately, not yet available.
Therefore, it is fair to say that by applying our results to
situations with relativistic velocities we do a 
somewhat blind extrapolation.
We think, however, that our results provide a correct order of magnitude
estimate of the real physical quantities. The same attitude was
adopted by the authors of previous works on the subject \cite{KKT,KMK}.

\section{Velocity and magnetic field spectrum for MHD
turbulence}\label{MHD}

Turbulence in an electrical conducting fluid in the presence of
magnetic fields is known as MHD turbulence. The dynamics of MHD
turbulent energy cascade may differ considerably from that of
conventional, Navier-Stokes (NS) turbulence, which we
discussed in the previous section. The main reason of such
different behaviour resides in the so called Alfv\'en effect
\cite{Biskamp}: in the presence of a strong background magnetic
fields ${\bf B}_0$ the velocity and magnetic field fluctuations
become strongly correlated, {\em i.e.}~${\bf v} \simeq \pm \, \delta {\bf B}/
{\sqrt {4\pi (\rho + p)}}$. Fluctuations of this kind correspond
to linearly polarized  Alfv\'en waves propagating along the field
${\bf B}_0$ with velocity
\be
\label{v_A}
 v_A = \frac{B_0}{\sqrt {4\pi (\rho + p)}}~,
\ee
the so called Alfv\'en velocity. 
In the absence of external magnetic fields ${\bf B}_0$ has to be intented as the
average field computed over a scale $L_0 \gg L = 2\pi/k$ . In our case we may
assume $L_0$ to coincide with $L_S = 2 \pi /k_S$.
MHD corrections to the turbulent cascade
can only be neglected if the turbulent velocity $u_k$ is larger
than $v_A$. In the opposite limit MHD turbulence sets-in which may
be figured as an ensemble of stochastic Alfv\'en waves.

Let's define $k_{\rm eq}$ as the wavenumber at which $u_k$ first
becomes equal to $v_A$. In the case $u_k$ is a decreasing function
of $k$, {\em i.e.}~$\gamma < 2/3$, we may assume a NS type spectrum for
$k < k_{\rm eq}$ and determine $k_{\rm eq}$ by equating
eq.~(\ref{u_k}) to (\ref{v_A}). It is evident that MHD turbulence
will never develop if $k_{\rm eq} > k_D$, where $k_D$ is the
damping momentum for the NS turbulence. In the opposite case,
however, there will be a range of momenta $k_D > k > k_{\rm eq}$
where we may expect noticeable corrections to the expressions
written is Sec.~\ref{turbo}. 
Indeed, since the Alfv\'en velocity
is larger than $u_k^{\rm NS}$ in this range, the time of interaction
of two eddies with similar momentum   $T_A \sim 1/k v_A$ is
smaller than the turnover time $T_k$. Since many interaction
events are necessary to change the wave packet amplitude
appreciably, the energy transfer time becomes longer
\cite{Biskamp}
\be
\label{MHDtime} T_k \rightarrow T ' _k \equiv T_A \left(\frac
{T_k}{T_A}\right)^2 = \frac {T_k^2}{T_A} \gg T_k ~.
\ee
This effect leads to  a milder turbulent spectrum. In the case of
stationary energy injection localized to a single momentum $k_S$,
eq.~(\ref{MHDtime}) implies
\begin{eqnarray}
  \varepsilon_k &\simeq& u_k^4 \frac{k}{v_A} \\
  E(k) &\simeq& (\varepsilon v_A)^{1/2}~k^{-3/2}
\end{eqnarray}
which is the so called Iroshnikov-Kraichnan spectrum. In the case
of a continuous external stirring power $P_{\rm ext}$ and $\gamma <
2/3$, ensuring a decreasing slope of the velocity $u_k$,
eq.~(\ref{E_di_k_3}) generalizes to
\be
\label{modiIK} E(k) \sim (\varepsilon_S v_A)^{1/2} k_S^{-\frac 3 4
\gamma}~k^{\frac 3 2\left(\frac \gamma 2 - 1\right)}~.
\ee
Since the magnetic fields at the GW emission time is not a direct
observable, it is convenient to express the previous results in
terms of  $k_{\rm eq}$ which may be directly
observed in the GW power spectrum.

From our previous considerations it follows that $k_{\rm eq}$ is
defined by
\be
 u^2(k_{\rm eq}) = \varepsilon_S^{2/3} k_S^{-\gamma}
 k_{\rm eq}^{\gamma - \frac 2 3} \equiv v_A^2~.
\ee
 By substituting this expression in eq.~(\ref{modiIK}) we get the result
\be
\label{generalE}
 E(k) \sim \left\{ \begin{array}{ccl}
  \varepsilon_S^{2/3} k_S^{-\gamma} k^{\gamma - 5/3} \; , & \; & k < k_{\rm eq}\\
  \varepsilon_S^{2/3} k_S^{-\gamma} k_{\rm eq}^{\frac \gamma 4 - \frac 1 6}~
  k^{\frac 3 2\left(\frac \gamma 2 - 1\right)} \; , & \;  & k > k_{\rm eq}	\; .
  \end{array}
  \right.
\ee

Since the GW power spectrum is directly related to $E(k)$ (see
section \ref{GWs}), we can already anticipate here that
possible observations of a variation in the the slope of the GW
power spectrum, fitting the power laws reported in
eq.~(\ref{generalE}), would provide information about the very
presence and the strength of a smooth (non-turbulent) magnetic
field component at the GW emission time.

Since Alfv\'en waves carry the same amount of magnetic and kinetic
energy, approximate equipartition between kinetic and magnetic
turbulent energies should hold for $k \geq k_{\rm eq}$. According
to the authors of Ref.~\cite{Durrer}, tangled magnetic fields
will also contribute to GW emission. Furthermore magnetic fields
may have produced observable effects on the
CMB anisotropies \cite{Durrer,CMB} and provided the seeds of galactic and
intergalactic magnetic fields \cite{report}. 
Therefore it is worthwhile to write
here also the magnetic field turbulent fluctuation spectrum. By
neglecting the turbulent magnetic energy above the equipartition
scale we have
\be
\label{magneticE} E_M(k) \sim \left\{ \begin{array}{ccl}
  0 \; , & \; & k < k_{\rm eq}\\
    \varepsilon_S^{2/3} k_S^{-\gamma} k_{\rm eq}^{\frac \gamma 4 - \frac 1 6}~
  k^{\frac 3 2\left(\frac \gamma 2 - 1\right)} \; , & \;  &  k > k_{\rm eq} \; ,
  \end{array}
  \right.
\ee
where $E_M (k)$ is the spectrum of energy stored in magnetic field fluctuations per unit
enthalpy of the fluid.

In order to cast the spectrum (\ref{generalE}) in a more symmetric form, we define
a new parameter $\delta$ and a new scale $k' _S$
\be \label{rescaling}
\delta = \frac{3}{4} \gamma+\frac{1}{6} \; , \qquad \qquad {k'_S}
^{-\delta} \equiv k_S ^{-\gamma} k_{\rm eq}^{\frac \gamma 4 - \frac
1 6}   \; ,
\ee 
so that eq.~(\ref{generalE}) becomes
 \be \label{generalEbis}
 E(k) \sim \left\{ \begin{array}{ccl}
  \varepsilon_S^{2/3} k_S^{-\gamma} k^{\gamma - 5/3} \; , & \; & k < k_{\rm eq}\\
  \varepsilon_S^{2/3} {k '_ S} ^{-\delta}
  k^{\delta - 5/3} \; , & \;  & k > k_{\rm eq} \; .
  \end{array}
  \right.
\ee
The same procedure with the same $\delta$ and $k'_S$ works for the relation between a turbulent
scale $k$ and the corresponding characteristic frequency $f_k$, because
$f_k \simeq \frac{1}{2 \pi} k^{3/2} E(k)^{1/2}$, see eqs.~(\ref{omega_k}) and (\ref{E_di_k_4}).
Thus we have
\be
f_k \simeq \left\{ \begin{array}{ccl}
\frac{1}{2 \pi} \: \varepsilon_S ^{1/3} \: k_S ^{-\gamma/2} \: k^{2/3+\gamma/2} \; , &
	\; & k < k_{\rm eq} \\
\frac{1}{2 \pi} \: \varepsilon_S ^{1/3} \: {k'_S} ^{-\delta/2} \: k^{2/3+\delta/2} \; , &
	\; & k > k_{\rm eq} \; .
  \end{array}
  \right.
\ee
In section \ref{GWs} we will compute the amount of relic GWs from NS turbulence.
The only specific features of NS turbulence that we will use
are the turbulent spectrum $E(k)$ and the relation between frequency
and length scale. With the above results in mind, this means that all the formulae of
section \ref{GWs} are straightforwardly applicable to MHD turbulence by the formal
replacements $\gamma \to \delta$ and $k_S \to k'_S$.

For $\gamma \geq  2/3$ MHD corrections will be relevant only if
$u_S < v_A$ \footnote{Clearly this will never be the case if $u_S$
is close to the speed of light.}. It is easy to verify that all
the previous expressions which were derived for $\gamma < 2/3$ can
be applied to the $\gamma \geq  2/3$ case by simply inverting their
range of validity with respect to $k$. Namely, we get a
MHD turbulent spectrum for $k < k_{\rm eq}$ and, if $k_{\rm eq} <
k_D$, a NS spectrum for $k > k_{\rm eq}$.

\section{Relic gravitational waves}\label{GWs}

We now want to compute the amount of GWs produced by cosmological
turbulence. We are interested in the present-day energy spectrum normalized to the
critical density,
\be
\Omega_{\rm gw} (f) \equiv \frac{1}{\rho_c} \frac{d \rho_{\rm gw}}{d \log f}    \; .
\ee
We will partly make use of the recent results of Kosowsky {\em et
al.}~\cite{KMK}: these were derived under the assumptions that the
turbulence has a Kolmogorov spectrum ($\gamma=0$, in the notations of section \ref{turbo})
and that it lasts for a time interval $\tau$ which is very short compared to the Hubble time
at that epoch.
We will perform a more general calculation, including the possibility that the plasma is stirred
at any scale and develops a general turbulent spectrum, as discussed in section \ref{turbo}.
We also consider the case that the time interval during which turbulence is active is long and
comparable to the Hubble time. We deal with this case assuming that the coherence time
of turbulence, which will be of the order of the characteristic turnover time
of a single turbulent eddy, is substantially smaller than the Hubble time, so that all the machinery
developed in the small $\tau$ approximation is applicable, because different eddies act as
uncoherent sources of GWs. The latter requirement is automatic if we consider eddies whose
characteristic length scale is well inside the horizon and whose
characteristic velocity is not too small.

Our convention for the Fourier transform of any function $F({\bf x})$ at fixed time is
\be
F ({\bf k}) \equiv \frac{1}{V} \int \! d^3 x \:
e^{+i {\bf k} \cdot {\bf x}} F ({\bf x})    \; ,
\ee
where ${\bf x}$ and ${\bf k}$ are physical quantities and we keep the
reference volume $V$ for dimensional convenience.

Let us consider a turbulent fluid in the Friedmann-Robertson-Walker (FRW) Universe at a 
fixed time $t_*$ and denote its velocity field as ${\bf u} ({\bf x})$.
The turbulent motions of the fluid will produce a stochastic background of GWs.
We define the perturbed metric as $g_{\mu\nu} = \hat g_{\mu\nu}(t) + h_{\mu\nu} ({\bf x}, t)$
where $\hat g_{\mu\nu}$ is the FRW unperturbed metric.
Under the assumptions that the fluid is incompressible and that turbulence is statistically
homogeneous and isotropic, the velocity correlator of the fluid can be parameterized as
\be
\langle \: u_i ({\bf k}, t) \: u_j ^* ({\bf k} ', t) \: \rangle =
\frac{(2 \pi)^3}{V} (\delta_{ij} - \hat k _i \hat k_j) \: \delta^3
({\bf k} - {\bf k} ') \: P(k)   .
\ee
It is straightforward to show that the velocity spectrum $P(k)$ is related to the
turbulent energy spectrum $E(k)$ defined in section \ref{turbo} by
\be \label{spettri}
E(k) \equiv \frac{1}{\rho+p} \frac{d \rho_{\rm turb}}{d k} = \frac{V}{\pi^2}
k^2 P(k)    \; .
\ee
If the source acts during a short time interval $\Delta t_*$ (see the discussion above),
much smaller than the Hubble time at $t_*$, we can neglect the expansion of the Universe, and use
the result of Kosowsky {\em et al.}~\cite{KMK} for the correlator of GWs produced during
the interval $\Delta t_*$,
\be \label{k_41}
\langle \: h_{ij} ({\bf k}, t_*) \: h_{ij} ^* ({\bf k} ', t_*)
\: \rangle \simeq
\frac{9 \sqrt{2} \: (16 \pi G)^2 \: \Delta t_* \: w^2}{16 \: k^3}
\delta^3 ({\bf k} - {\bf k} ') \int \! d^3 q \: P(q) P(|{\bf k} - {\bf q}|) \; ,
\ee
where $w \equiv \rho + p$ is the enthalpy density of the fluid \footnote{
The r.h.s.~of eq.~(\ref{k_41}) is 4 times larger than the corresponding quantity appearing
in ref.~\cite{KMK}. This is because our definition of the metric perturbation $h_{ij}$
is twice larger than the one used in that paper, where $h_{ij} \equiv \frac{1}{2} \delta g_{ij}$.
Although the two descriptions are equivalent, one must pay attention when applying them to
calculate the energy density of the GW background. In ref.~\cite{KMK} this is done by
means of formulae taken from ref.~\cite{maggiore}, which instead uses our definition of
$h_{ij}$. Therefore we think that the authors of ref.~\cite{KMK} underestimates the 
GW energy density by a factor of 4.  
}.

For a power-law spectrum $P(k)= A \: k^{-\alpha}$ in the range
$k_S < k < k_D$, the integral appearing in eq.~(\ref{k_41}) is well approximated by
\cite{KMK}
\be
\int \! d^3 q \: P(q) P(|{\bf k} - {\bf q}|) \simeq         \label{integral}
4 \pi A^2 \left[
- \frac{k^{3 -2 \alpha } \: \alpha}{(3 -\alpha ) (3 -2\alpha)}
+ \frac{k_D^{3-2\alpha}}{3 -2\alpha} -
\frac{k^{-\alpha} k_S ^{3-\alpha}}{(3-\alpha )}
\right]
\ee
for $k_S < k < k_D$ and 0 outside that interval. Eq.~(\ref{integral}) holds for $\alpha > 3/2$:
for $\alpha \le 3/2$ the integral must be computed more carefully, but this case will not be
of interest for us.
There are three qualitatively different regimes of eq.~(\ref{integral}): for $\alpha>3$ the last
term of the r.h.s.~is dominant, while for $\alpha<3$ the first dominates. The limiting case
$\alpha \to 3$ gives a term which behaves like $k^{-3} \log k$, coming from the sum of the first
and the last term. Taking into account the relations
$\alpha = 11/3 - \gamma$ and $A \: V/\pi^2= u_S ^2 \: k_S^{2/3 -\gamma}$, coming from comparing
eqs.~(\ref{spettri}) and (\ref{E_di_k_4}), we obtain
\be \label{casi}
\int \! d^3 q \: P(q) P(|{\bf k} - {\bf q}|) \simeq \frac{4 \pi^5}{V^2}
\: u_S ^4 \: k_S^{4/3 -2\gamma} \times
\left\{ \begin{array}{ccl}
\frac{3}{2-3\gamma} \: k_S ^{-2/3+\gamma} \: k^{-11/3+\gamma} \; ,& \; & \gamma<2/3 \\
\frac{3(11-3\gamma)}{(3\gamma-2)(13-6\gamma)} \: k^{-13/3+2\gamma}\; , &  \; & \gamma>2/3 \\
k^{-3} \log \frac{k}{k_S}    \; ,&   \; & \gamma \to 2/3
    \end{array} \right.
\ee
as leading contributions.

It is now straightforward to compute the energy spectrum of
gravitational waves $h^2 _0 \Omega_{\rm gw} (f)$ for the three different cases of
eq.~(\ref{casi}). In the Appendix we present all the details of the computation.
Here we summarize only the strategy and the results.
From eq.~(\ref{k_41}) one computes the real-space correlation function
\be
\langle \: h_{ij} ({\bf x}, t_*) \: h_{ij} ({\bf x}, t_*)
\: \rangle =
\frac{V^2}{(2\pi)^6} \int \! d^3k d^3 k' e^{i({\bf k'}-{\bf k})\cdot {\bf x}}
\langle \: h_{ij} ({\bf k}, t_*) \: h_{ij} ^* ({\bf k} ', t_*)
\: \rangle  \; ,
\ee
and from that the {\em characteristic amplitude} $h_c(f)$ of GWs at frequency $f$,
defined by (see ref.~\cite{maggiore} for details)
\be
\langle \: h_{ij} ({\bf x}, t) \: h_{ij} ({\bf x}, t)
\: \rangle \equiv
2 \int_0 ^\infty \frac{df}{f} h^2 _c (f,t)  \; ,
\ee
where one has to express everything in terms of the frequency $f$ of a turbulent scale
$k$ by means of the relation
\be \label{freq}
f_k = \frac{\omega_k}{2\pi} \simeq \frac{1}{2\pi} \: u_S \:
k_S^{1/3 - \gamma/2} \: k^{2/3 + \gamma/2}
\ee
following from eq.~(\ref{omega_k}).

The expansion of the Universe redshifts the frequency and damps the amplitude of GWs: both
are inversely proportional to the scale factor $a$. If the subscripts 0 and $*$
stand respectively for the present time and the time of production,
we have \footnote{Although we use here a parameterization suitable to
  describe GW production around neutrino decoupling time, we should
  stress that the expressions contained in this section are completely general.}
\be
\frac{a_*}{a_0}  \simeq  1.7 \times 10^{-10}
        \left[ \frac{10.75}{g_*} \right] ^{1/3}
        \left[ \frac{\rm MeV}{T_*} \right]      \; ;
\ee
the frequency we observe today is thus
\be
f_0 \simeq 1.15 \times 10^{-7} \: {\rm mHz} \: \left[ \frac{f_*}{H_*} \right]
\left[ \frac{T_*}{\rm MeV} \right] \left[ \frac{g_*}{10.75} \right] ^{1/6}
\label{frequency}       \; ,
\ee
where we used the fact that the Hubble parameter at time of production is
\be
H_*  =  1.66 \: g_* ^{1/2} \: \frac{ T_* ^2}{M_P} \simeq 4.5 \times 10^{-22} \: {\rm MeV}
    \left[ \frac{T_*}{\rm MeV} \right]^2
    \left[ \frac{g_*}{10.75} \right] ^{1/2}
\ee
and that $1 \: {\rm mHz}  \simeq  6.7 \times 10^{-25} \: {\rm MeV}$.
From eq.~(\ref{frequency}) we see that in order to have detectable ({\em i.e.}~high enough:
for instance LISA operates at $f \sim$ mHz) frequencies we must have turbulent
motions at length (time) scales well below the horizon and at large temperatures $T_*
\simgeq 1$ MeV. For such a reason we will be interested in the high-frequency region of
the spectrum: the maximum frequency we can expect today, in terms of the damping scale
$k_D$ at time of production, is
\be \label{cutoff}
f_{\rm max} \equiv \frac{a_*}{a_0} f_{k_D} \simeq 2 \times 10^{-8} \: {\rm mHz}
\left( \frac{k_D}{H_*} \right)^{2/3+\gamma/2}
u_S
\left[ \frac{k_S}{H_*} \right]^{1/3-\gamma/2}
\left[ \frac{T_*}{\rm MeV} \right] \left[ \frac{g_*}{10.75} \right] ^{1/6}  \; .
\ee

The characteristic amplitude measured today at a frequency $f$ is
\be     \label{red}
h_c (f,t_0) = \frac{a_*}{a_0} h_c ( f_* = \frac{a_0}{a_*} f , t_*)  \; ,
\ee
and the energy spectrum we are interested in is related to that by \cite{maggiore}
\be \label{relaz}
h_c (f) = 1.3 \times 10^{-15} \left(\frac{\rm mHz}{f} \right)
\sqrt{h_0 ^2 \Omega_{\rm gw}(f)} \; .
\ee
Applying this procedure to the three different cases of eq.~(\ref{casi}) leads to (see the Appendix
for the details)
\begin{eqnarray}
\gamma<2/3 \quad \Rightarrow \quad
        h_0 ^2 \Omega_{\rm gw} (f) & \simeq & 6 \times 10^{-4}       \label{omega_min}
        \:
        \frac{\Delta t_*}{H_* ^{-1}}
        u_S ^6
        \left[ \frac{k_S}{H_*} \right]^{-1}
        \left[ \frac{g_*}{10.75} \right] ^{-1/3}
        \: C_\gamma \: 
        \left( \frac{f}{f_S} \right)^{-\frac{2(7-6\gamma)}{4+3\gamma}}
        \\
\gamma > 2/3 \quad \Rightarrow \quad
        h_0 ^2 \Omega_{\rm gw} (f) & \simeq &  6 \times 10^{-4}      \label{omega_mag}
        \:
        \frac{\Delta t_*}{H_* ^{-1}}
        u_S ^6
        \left[ \frac{k_S}{H_*} \right]^{-1}
        \left[ \frac{g_*}{10.75} \right] ^{-1/3}
        \: D_\gamma \: 
        \left( \frac{f}{f_S} \right)^{-\frac{18(1-\gamma)}{4+3\gamma}}
        \\
\gamma \simeq 2/3 \quad \Rightarrow \quad
        h_0 ^2 \Omega_{\rm gw} (f) & \simeq &  4 \times 10^{-5}             \label{omega_ug}
        \:
        \frac{\Delta t_*}{H_* ^{-1}}
        u_S ^6
        \left[ \frac{k_S}{H_*} \right]^{-1}
        \left[ \frac{g_*}{10.75} \right] ^{-1/3}
        \: 
        \left( \frac{f}{f_S} \right)^{-1} \log \frac{f}{f_S}
    \; ,
\end{eqnarray}
where $C_\gamma = \frac{1}{(2-3\gamma)(4+3\gamma)}$ and 
$D _\gamma = \frac{(11-3\gamma)}{(3\gamma-2)(4+3\gamma)(13-6\gamma)}$ are ${\cal{O}}(0.1)$ 
numerical coefficients and $f_S$ is the properly redshifted frequency corresponding today 
to the largest length scale at time of production,
\be \label{fS}
f_S  \simeq 2 \times 10^{-8} \:
u_S
\frac{k_S}{H_*}
\left[ \frac{T_*}{\rm MeV} \right] \left[ \frac{g_*}{10.75} \right] ^{1/6}  \: {\rm mHz}\; .
\ee

In the Kolmogorov case we get
\be
\label{KolmogorovGW}
\gamma=0 \quad \Rightarrow \quad
h_0 ^2 \Omega_{\rm gw} (f) \simeq
6 \times 10^{-32}
\:
\frac{\Delta t_*}{H_* ^{-1}}
\left( \frac{\rm mHz}{f} \right)^{7/2}
u_S ^{19/2}
\left[ \frac{k_S}{H_*}  \right]^{5/2}
\left[ \frac{T_*}{\rm MeV} \right]^{7/2}
\left[ \frac{g_*}{10.75} \right] ^{1/4}
\; .
\ee
Apart for the factor 4 which we already discussed, this expression is consistent with the 
result of KMK \cite{KMK}.
It is evident from eq.~(\ref{KolmogorovGW}) that a Kolmogorov turbulent spectrum
give rise to an undetectable GW intensity at mHz frequencies if $T \sim 1~{\rm MeV}$ and 
the energy injection scale is comparable to the Hubble horizon size.
However, as we will see in more details below, observationally more promising 
intensities can be obtained, even for a Kolmogorov spectrum, for larger emission 
temperatures especially in those case in which the injection momentum $k_S$ increases more 
rapidly than $H_*$ with $T_*$.

For $\gamma>2/3$ the dependence of $h_0 ^2 \Omega_{\rm gw}$ on $f$ is milder than in the
$\gamma<2/3$ case: for $\gamma > 1$ the spectrum is even increasing with $f$.
For $\gamma=1$ the spectrum is flat,
\be
\gamma=1 \quad \Rightarrow \quad
h_0 ^2 \Omega_{\rm gw} \simeq 10^{-4} \:
\frac{\Delta t_*}{H_* ^{-1}}
\: u_S ^6
\left[ \frac{H_*}{k_S} \right]
\left[ \frac{10.75}{g_*} \right] ^{1/3}     \; .
\ee

\subsection{Time-integration}
If the time during which turbulence is active is long, {\em i.e.}~of the order of the Hubble time
$H_* ^{-1}$, we have to perform an integration over production time $t_*$ in order to get the correct
$h_0^2 \Omega_{\rm gw}$.
To be more specific, let's consider the case of GWs produced at the time
of neutrino decoupling which we will illustate in more details in the
next section.
Consider for simplicity a turbulent spectrum saturated at relativistic velocities $u_k \sim 1$:
as discussed in section \ref{turbo}, it is well described by the $\gamma=2/3$, $u_S =1$ case.
The energy density of GWs produced in a short time interval $d t_*$ is given by
eq.~(\ref{omega_ug}) with the $u_S$ factor set to 1.
Neglecting for simplicity the logarithmic factor (this a conservative assumption
since $f>f_S$) and using eq.~(\ref{fS}) we obtain
\begin{eqnarray}
h_0 ^2 \Omega_{\rm gw} (f) & \simeq &  8 \times 10^{-13}
\int _{t_{\rm on}} ^{t_{\rm off}} \!
\frac{d t _*}{H_* ^{-1}}
\left( \frac{\rm mHz}{f} \right)
\left[ \frac{T_*}{\rm MeV} \right]
\left[ \frac{10.75}{g_*} \right] ^{1/6}     \nonumber \\
& \simeq &  8 \times 10^{-13}
\left( \frac{\rm mHz}{f} \right)
\left[ \frac{10.75}{g_*} \right] ^{1/6}
\int _{T_{\rm off}} ^{T_{\rm on}} \!
d \left( {T_*}/{\rm MeV} \right)        \nonumber \\
& \simeq &
8 \times 10^{-13}
\left( \frac{\rm mHz}{f} \right)
\left[ \frac{T_{\rm on}}{\rm MeV} \right]
\left[ \frac{10.75}{g_*} \right] ^{1/6}     \; ,
\end{eqnarray}
where we assumed $T_{\rm on} \gg T_{\rm off}$ and $g_* \simeq \rm const$.

The cutoff frequency is (see eq.~(\ref{cutoff}))
\be \label{cut_neutrini}
f_{\rm max} \simeq 2 \times 10^{-8} \: {\rm mHz}
\left( \frac{k_D}{H_*} \right)
\left[ \frac{T_*}{\rm MeV} \right]
\left[ \frac{g_*}{10.75} \right] ^{1/6}
\; .
\ee
For the relativistic case $u_k \sim 1$ the damping scale is $k_D \sim 1/\eta$,
see section \ref{turbo}; in the case discussed in section \ref{neutrino} the viscosity
is $\eta = \frac{4 \: \rho_\nu}{15 (\rho+p)} \: \ell_\nu \simeq \frac{1}{30} \ell_\nu$. The
mean free path $\ell_\nu$ scales like $T^{-5}$, and at neutrino decoupling it is equal to
the horizon length. This gives
\be
\label{fmax}
f_{\rm max} \sim 1 \: {\rm mHz}
\left[ \frac{T_{\rm on}}{40 \: \rm MeV} \right]^4
\left[ \frac{g_*}{10.75} \right] ^{1/6}
\; ,
\ee
where we used $T_d \simeq$~1 MeV.

\section{GWs from turbulence produced by neutrino inhomogeneous diffusion}\label{neutrino}

In Ref.~\cite{dario} two of us showed that if isocurvature
fluctuations existed in the early universe under the form of a
space dependent neutrino net number ($N_a({\bf x}) \equiv
n_{\nu_a}({\bf x}) - n_{\bar \nu_a}({\bf x})$,\ \ $a =
e,\mu,\tau$) turbulence should have developed in the primordial
plasma before neutrino decoupling. Such an effect arises as a
consequence of  neutrino currents which flow along the lepton
number gradients when the neutrino mean free path $\ell_\nu(T)$
becomes comparable to the characteristic size $\lambda $ of the
isocurvature fluctuation. The residual elastic scattering of the
diffusing neutrinos onto electrons and positrons will then
accelerate these particles together with the photons and the rest
of the neutrinos to which $e^-$ and $e^+$ are still tightly
coupled (baryons can be disregarded at that time). A random
distribution of fluctuations will generally give rise to vortical
motion of this composite fluid. Depending on the velocity and the
size of the eddies, turbulence may then develop in the interval of
time  during which the random forces due to neutrino elastic
scattering overcome the shear viscosity force \footnote{This is
the same of saying that the Reynolds number is much larger than
one.}. It was showed that the fluid velocity can approach the
speed of light if the amplitude of the fluctuation $\delta N/N$ is
not too much smaller than 1. In such a  case the authors of
\cite{dario} claimed that turbulence should give rise to a
stochastic background of GWs at a level which may be detectable by
LISA space observatory or by its upgrading. In this section we
investigate this issue in more details.

The fluid acceleration mechanism is suitably described by kinetic
theory in an expanding geometry \cite{Bernstein}. The first step
of the computation is to determine the neutrino momentum flux by
solving the transport equation for the neutrinos in the presence
of a source term given by the fluctuation of $N_a$ and a collision
term due to neutrino-electron(positron) scattering.

Let's first introduce some useful notation:
$x \equiv \frac{t}{t_d}$ is time in units of
	the neutrino decoupling time $t_d$;
$H(t) = \frac{1}{2 t_d} \:  x^{-1} $  and
	$H_d = \frac{1}{2 t_d}$ are the Hubble rates for
	$t_{QCD} < t < t_d$   and at $t = t_d$;
$\lambda (t) = 2 t_d \: \tilde\lambda \: x^{1/2} $ and
	$\tilde \lambda \equiv \lambda(t_d) \: H_d$ is the fluctuation
 	wavelength in natural units and in units of $H_d^{-1}$;
$k(t) = \frac{1}{2 t_d} \: \tilde k \: x^{-1/2}$ and
 	$\tilde k \equiv k(t_d) / H_d$  are  the corresponding wavenumbers;
$\ell_\nu (t) = \tau = 2 t_d \: x^{5/2}$  are the neutrino mean
 	free path and collision time;
${\bf K}_a({\bf x}, t) \equiv \frac 1 \rho_{\nu_a} \int {\bf k} \:  f_{\nu_a}(E,
 	{\bf k}) \frac {d^3 k}{(2\pi)^3}$  is the specific
 	momentum flux of $\nu_a$;
finally, we define the derivative $(\:\:)' \equiv \frac{\pa}{\pa x} (\:\:) 
	=t_d \frac{\pa}{\pa t} (\:\:)$.
Then the momentum transport equation can be written \cite{dario}
\be \label{transport}
 \frac{\pa}{\pa t} {\bf K}_a \simeq
 -\frac{1}{3}{\bf \nabla}
    \left( \frac{\delta N_a}{N_a} \right) 
    -4 H \: {\bf K}_a - \frac{2}{\tau}
     \left( {\bf K}_a -{\bf v} \right)   \; ,
\ee
where ${\bf v}$ is the macroscopic velocity of the $e^+ e^- \gamma$ fluid.
The Euler equation for ${\bf v}$ is
\be \label{Euler}
\frac{\pa}{\pa t} {\bf v} \simeq  \frac{4}{\tau}
    \: \frac{\rho_\nu}{ \gamma^2 (\rho+p)}
    \: \left( {\bf K}_a - \bf v \right)
    -H {\bf v} + \eta \nabla^2 \left({\bf v}  - {\bf K}_a\right)    \; ,
\ee
where $\eta = \frac{4 \: \rho_\nu}{15 (\rho+p)} \: \ell_\nu$
is the shear viscosity due to neutrino diffusion and we assumed for simplicity 
to have the same inhomogeneities in the density of $\nu$'s and $\bar \nu$'s, 
{\em i.e.}~${\bf K}_\nu \simeq {\bf K}_{\bar \nu}$ 
{\footnote{As long as we are not concerned about magnetic
field production (see below), we can safely approximate the
$\nu-e^-$ and $\nu-e^+$ mean collision time to be equal. The
slight differences between the values of $\ell_\nu$ and $\tau$ for
the $\nu_e$ and the $\nu_{\mu,\tau}$, are also not significant
here.}}.
Finally, the evolution of the fluctuations $\delta N_a/N_a$ is governed
by the continuity equation
\be
\frac{\partial}{\partial t} N_a = - \nabla \cdot \left( N_a {\bf K}_a \right)     \; .
\ee

By making explicit the $x$ dependences and passing to the Fourier space we transform 
the last three coupled equations into
\be     \label{system}
\left\{
\begin{array}{l}
{K_k}' = -\frac{i}{6} \: \tilde k \:
        \delta _k \: x^{-1/2} -  2 x^{-1} \:K_k - x^{-5/2} \left( K_k-v_k \right)               \\
v_k ' = 2 r \: x^{-5/2}\:  (K_k - v_k) - \frac{1}{2} x^{-1} \: v_k - \frac{2 r}{15}
    \: \tilde k ^2 \: x^{3/2} \left( v_k-K_k \right)    \\
\delta_k '=-\frac{i}{2} x^{-1/2} \: \tilde k \: K_k
\end{array}
\right.
\ee
where $\delta _k$ is the Fourier transform of
$\left( \frac{\delta N_\nu}{N_\nu} \right)$
and we defined $r \equiv \frac{\rho_\nu}{\rho+p} \:$.
In eq.~(\ref{system}) we kept only the first order
in $\delta_k$ and we ignored the Lorentz factor $\gamma$ of eq.~(\ref{Euler})
\footnote{The sound velocity $1/ \sqrt{3}$
sets an upper bound for macroscopic velocities of the fluid,
therefore $1 < \gamma < \sqrt{3/2} \simeq 1.2$.}. 
The initial conditions at $x=0$ are $K_k = v_k = 0$, $\delta_k = \delta_k ^0$.

The characteristic decay time $\bar x$ of the fluctuation $\delta_k$ can be calculated
by exactly solving the system (\ref{system}). For $x \ll \bar x$ the fluctuation can be
considered frozen at its initial value $\delta_k ^0$; at that early time the system (\ref{system})
reduces to
\be     \label{system_approx}
\left\{
\begin{array}{l}
{K_k}' = -\frac{i}{6} \: \tilde k \:
        \delta _k ^0 \: x^{-1/2} -  2 x^{-1} \:K_k - x^{-5/2} \left( K_k-v_k \right)            \\
v_k ' = 2 r \: x^{-5/2}\:  (K_k - v_k) - \frac{1}{2} x^{-1} \: v_k      \; , 
\end{array}
\right.
\ee
where we kept the leading order in $1/x$.
The system (\ref{system_approx}) is easily diagonalized and approximately solved by
\be
K_k (x\ll \bar x) \simeq v_k (x \ll \bar x) \simeq  -\frac{i}{6} \: \tilde k \:
        \delta _k ^0 \: \frac{2 r}{1+5 r} x^{1/2}       \; ,
\ee
with $K_k-v_k = {\cal{O}}(x^2)$. Plugging this into the continuity equation for $\delta_k$ gives
$\bar x \simeq \frac{6(1+5 r)}{r \: \tilde k ^2}$ as characteristic decay time.
At later times $x \simgeq \bar x$ the source term is negligible and the neutrino 
momentum $K_k$ and the fluid velocity $v_k$ are damped by the expansion of the Universe (the viscosity
term $\sim \tilde k ^2 \; x^{3/2} \left( v_k-K_k \right)$ will become important even later). Therefore
the $e^+ e^- \gamma$ fluid reaches the maximum velocity at time $x \sim \bar x$,
\be     \label{v_max}
v _k ^{\rm max} \sim v_k (\bar x) \sim \delta ^0 _k  \sqrt{\frac{2 r}{3(1+5 r)}}                \; , 
\ee
which has the same $k$ dependence as the initial fluctuation $\delta ^0 _k$.
This rough analytic estimate is confirmed and made more precise by exact numerical 
integration  of the coupled equations (\ref{system}): eq.~(\ref{v_max}) overestimates the actual value
of $v _k ^{\rm max}$ by a factor $1.2 - 1.3$.

As an illustration, let's first consider GW production in the simple case of isocurvature 
fluctuations peaked at a single comoving wavenumber ${\tilde k}_S$. 
When $\ell_{\nu}$ becomes comparable to $k_S$, let's call this time $t_*$, neutrino 
diffusion give rise to a vortical velocity field according
to the mechanism discussed in the above. Let's assume for simplicity that the peak velocity 
$v_{k_S} \equiv u_S$ is reached instantaneously. This velocity will be maintained for a 
time of the order of the Hubble time $H_*^{-1} \simeq t_*$. Since, the velocity turnover 
time is of the order $k_S^{-1}$, which is supposed to be much smaller than 
$H_*^{-1}$, turbulent energy cascade from $k_S$ down to $k_D$ have all the time to develop. 
This is not enough, however, to conclude that fully developed turbulence is actually formed.
In order to verify if this is the case we have to verify if the Reynolds number is much 
larger than unity.  We have
\be
{\rm Re} = \frac{u_S k_S^{-1}}{\eta} \simeq 30 \: u_S
\ee
where we used the expression $\eta = \frac{4 \: \rho_\nu}{15 (\rho+p)} 
\: \ell_\nu \simeq \frac{1}{30} \ell_\nu $ for the viscosity and 
we assumed $k_S^{-1} \sim \ell_\nu$. 
For a localized stirring source the standard theory of turbulence would not
hold for such a low Reynolds number. This is because the required statistical isotropy
and homogeneity of the turbulent velocity field cannot be achieved in this case.
The situation may however be different if the stirring source is stochastic itself 
providing statistical isotropy and homogeneity from the very beginning.
We assume that in such a situation the results of Secs.\ref{turbo} and \ref{MHD} can be 
applied also for order unity Reynolds numbers, provided that $k_D$ is larger than $k_S$. 
Furthermore, as follows from eq.~(\ref{kD}), $k_D$ increases with respect to $k_S$ when 
$\gamma > 0$. The most favourable case is that in which isocurvature fluctuations give rise 
to fluid velocity close to unity over a wide range of wavenumbers.    
As we discussed in Sections \ref{turbo} and \ref{GWs}, this give rise to a saturated 
turbulent spectrum with $\gamma = 2/3$. In such a case eq.~(\ref{omega_ug}) holds for the 
GW energy density. 

In general the turbulent velocity power spectrum will depends on the power spectrum 
of the neutrino number fluctuations which in turn will depend on the 
mechanism responsible for their generation.
The computation of such spectrum  is beyond the aim of this paper. 
In Ref.~\cite{dario} the authors considered two possible scenarios. 
In the first one $N_{\nu}$ fluctuations with amplitude of order 1 are produced
as a consequence of active-sterile neutrino oscillations according to the
mechanism proposed by Di Bari \cite{DiBari} (see also Ref.~\cite{dolrep2}). 
The seeds of the {\em neutrino domains} are tiny fluctuations in the baryon 
number which may have been produced during the QCD phase transition or the inflation.
Their power spectrum will determine that of neutrino domain hence also that of GWs.
A continuous spectrum is generally to be expected.
Although in principle this mechanism may naturally give rise to very large amplitude
isocurvature fluctuations and GW signal, unfortunately it would be probably
hardly detectable by LISA. This is a consequence of the low critical temperature at which 
neutrino domains may have been formed which turns into a low GW frequency
(see eq.~(\ref{fmax})).
Better observational perspectives are offered by the second mechanism considered 
in \cite{dario} which is based on a generalization of the Affleck-Dine baryogenesis
mechanism \cite{DolKir,dolrep1}. In this case neutrino domains may be formed 
during inflation when a scalar field carrying lepton number rolls along (nearly) flat
directions of the superpotential. Isocurvature fluctuations of amplitude 
as large as 1 may be formed without invoking too extreme assumptions.
Although the fluctuations power spectrum is  expected to be strongly model dependent,
a nearly flat spectrum is a quite reasonable possibility. 
As we discussed above, a flat spectrum of the neutrino number fluctuations 
of amplitude close to 1 should give rise to a GW signal as given by eq.~(\ref{omega_ug})
which may be detectable by LISA.
Although this model does not provide an existence proof of 
GWs produced by neutrino inhomogeneous diffusion it gives, in our opinion, 
at least a reasonable plausibility argument.

\subsection{MHD effects}
  
In Ref.~\cite{dario} it was showed that magnetic fields should be
produced during the fluid acceleration process. This is a
consequence of parity violation in the standard model which turns
into a difference between the $\nu_a e^-$ and $\nu_a e^+$ cross
sections. It follows that an electric current ${\bf J}_{\rm ext}$
appears in the $e^+~e^-$ plasma in the presence of a net flux of
neutrinos. Although the magnetic field produced by this current is
initially very small, the coherent motion of the fluid induced by
the neutrino flow amplifies this seed exponentially with time
until equipartition is reached between the magnetic and fluid
kinetic energies. Numerical simulations \cite{dario} show that in
the range of parameters which is interesting from the point of
view of GW detection, equipartition is reached well before fluid
motion is damped by the viscosity and the universe expansion. It
is clear that in such a situation the effects of the magnetic
field on turbulence development have to be taken into proper
account.

The results we derived in Sec.~\ref{MHD} find here a natural
application. Let us first observe that, since equipartition is
reached already during the stirring phase, the equality $u_S
\simeq v_A$ holds in this case. As we discussed in Sec.~\ref{MHD},
depending on the characteristic of the energy injection spectrum
$P_{\rm ext}(k)$ two cases have to be distinguished. If $P_{\rm
ext}(k)$ is a decreasing function of $k$ ($\gamma < 2/3$), or it is
 a delta-function peaked at $k = k_S$ ($\gamma = 0$, as for
Kolmogorov type turbulence), the turbulent velocity will be
always smaller than the Alfv\`en velocity. This means that MHD
corrections can never be disregarded and the second of
eqs.~(\ref{generalE}) ($k > k_{\rm eq}$) \footnote{Note that
$k_{\rm eq} = k_S$ in this case.} applies for the turbulent
velocity spectrum. If, however, $P_{\rm ext}(k)$ is growing with
$k$ ($\gamma > 2/3$), or it is a constant ($\gamma = 2/3$, this is
the case for a saturated spectrum), we always have $u_k \geq v_A$ and
MHD effects are less important.

\section{GWs from turbulence produced by a first-order phase transition}\label{EWPT}

A first order phase transition proceeds through the nucleation of bubbles of
the true-vacuum phase inside the false vacuum. Once a bubble is nucleated, if its radius
is larger than a critical value, it begins to expand at velocity $v_b$, which for very
strong phase transitions can approach the speed of light \cite{KKT}.
Once the bubble walls begin to collide, they break spherical symmetry and thus
stir up the plasma at a scale comparable with their radii at the collision time. If the
Reynolds number of the plasma is high enough, the energy released in coherent motions
of the fluid is transferred to smaller scales and a turbulent spectrum establishes.
It must be stressed that the stirring process must last for an enough long time in order to
give rise to a fully developed turbulence. If this is not the case, as long as we are interested
in GW production we can treat the resulting partially developed turbulence as if it 
were a fully developed one persisting for a time interval of the order of the turnover
time on the largest stirring scale \cite{KMK}.

The rate of nucleation of one critical bubble per unit volume, $\Gamma$, 
is suppressed by the exponential of the bubble Euclidean action. As
the temperature of the Universe decreases the rate $\Gamma$ gets larger and larger; 
when at time $t_*=0$ the rate becomes comparable to $H^4$ the transition begins. 
In a neighbourhood of that time one can expand the Euclidean action, thus getting \cite{TWW}
\be
\Gamma \sim H_*^4 \: e^{\beta  t}       \; ;
\ee
$\beta^{-1}$ sets the time scale of the process: in a time interval of order 
$\beta^{-1}$ the transition is complete, {\em i.e.}~all the volume has been converted 
to the true vacuum phase. The characteristic length scale of the process is thus $v_b \: \beta^{-1}$.
As pointed out in ref.~\cite{TWW}, in realistic cases
$\beta$ is much larger than $H_*$, so that cosmic expansion can be neglected for the whole
duration of the process.

We now want to understand what turbulent spectrum one can expect
from such a first order phase transition. At generic time $t$ the distribution of bubbles
with a given radius $R$ is \cite{TWW}
\be
\frac{d N}{d R} (t) = \frac{1}{v_b} \: \Gamma(t_R) \: p(t_R)    \; ,
\ee
where $p(t)$ is the probability that at time $t$ a random point is in the false vacuum state
and $t_R = t-R/v_b$ is the time of nucleation of a bubble that at time $t$ has radius $R$ (the radius
at time of nucleation is completely negligible). The factor $p(t_R)$ suppresses the bubble
distribution at small radii, because smaller bubbles were nucleated at later times, when
the fraction of volume still in false vacuum phase was smaller. We will show that collisions 
of small bubbles give a negligible contribution to the stirring of the plasma, so that from now on
we will conservatively set $p(t_R) = 1$.

The energy carried by the expansion of a bubble of radius $R$ is proportional 
to its volume $R^3$. The energy distribution of bubbles at time $t$ is thus
$\frac{d E}{d R} (t) \propto e^{\beta t} \: R^3 \: e^{-\beta R / v_b}$,
which in momentum space reads
\be     \label{energy}
\frac{d E}{d k} (t) \propto e^{\beta t} \: k^{-5} \: e^{-\frac{2\pi \beta}{v_b k}}      \; ,
\ee
where we set $k = 2 \pi/R$. 
The plasma stirring process takes place when two (or more) bubbles collide: we expect that a
collision will release energy in bulk motions at a scale comparable to the radius
of the smaller colliding bubble, and the amount of energy released will be approximately
proportional to its energy. Since the energy distribution of small bubbles at any time $t$ decreases
like $k^{-5}$, from eq.~(\ref{E_di_k}) we expect to have no substantial deviation at large $k$
from Kolmogorov spectrum when turbulence develops. On the other hand, in principle we
can expect deviation from Kolmogorov spectrum at scales $k \sim \frac{2\pi \: \beta}{v_b}$,
{\em i.e.}~at the characteristic scale of the transition.
In order to understand if this is the case we must estimate the width of the stirring spectrum
$P_{\rm ext} (k)$: it is reasonable to assume that to be of the same order of magnitude
of the width of the bubble energy distribution (\ref{energy}). The latter is peaked at 
\be     \label{kS}
k_S = \frac{2 \pi \beta}{5 v_b} 
\ee
and has a width 
$\Delta k \simeq 1.4 \times k_S $, if we define the width as
the range of momenta $k$ out of which the energy distribution $d E / d k$ gets smaller than $1 / e$
times its maximum value $d E / d k |_{k_S}$.

The width $\Delta k$ is small, {\em i.e.}~most
of the energy of the transition is carried by bubbles of characteristic scale very close to $k_S$.
This means that the stirring spectrum $P_{\rm ext} (k)$ has a width which is below (or comparable to) 
the precision of the results on turbulent motions, which are usually derived under the uncertainty
condition $\Delta k/k \sim 1$ (see section \ref{turbo}).
Therefore, we do not expect any substantial deviation from Kolmogorov spectrum 
in the case of turbulence produced by a first order phase transition, {\em i.e.}~we can set $\gamma=0$ in
all the formulae of section \ref{GWs}. 

It is interesting to deal with the electroweak phase transition (EWPT), because it takes place at the
temperature $T_* \sim 100$ GeV and therefore, if it is of the first order, it gives rise
to a GW background peaked at a the frequency (see 
eqs.~(\ref{fS}) and (\ref{kS}))
\be     \label{fS_EWPT}
f_S  \simeq 3 \times 10^{-3} \:
u_S \: v_b^{-1}
\frac{\beta}{H_*}
\left[ \frac{T_*}{100 \: \rm GeV} \right] \left[ \frac{g_*}{100} \right] ^{1/6}  \: {\rm mHz}\; ,
\ee
which is in the range of maximum sensitivity of LISA if $u_S$ is relativistic and 
$\beta/H_* \sim 100$.
As pointed out in ref.~\cite{KMK} the duration of the turbulent source of GWs is roughly 
the turnover time on the largest length scale, $\Delta t_* \sim \beta^{-1} v_b / u_S$. 
The amount of relic GWs at milliHertz frequencies is thus (see eq.~(\ref{KolmogorovGW}))
\be     \label{omega_bubble}
h_0 ^2 \Omega_{\rm gw} (f) \simeq
6 \times 10^{-14}
\:
\left( \frac{\rm mHz}{f} \right)^{7/2}
u_S ^{17/2} \: v_b ^{-3/2}
\left[ \frac{\beta}{H_*}  \right]^{3/2}
\left[ \frac{T_*}{100 \: \rm GeV} \right]^{7/2}
\left[ \frac{g_*}{100} \right] ^{1/4}
\; .
\ee

If the bubbles expand as detonation fronts, {\em i.e.}~faster than sound,
the expansion velocity of the bubbles, $v_b$, and the turbulent velocity on the largest scale, $u_S$,
can be derived in terms of $\alpha$, the ratio between the false vacuum energy and the
plasma thermal energy at the transition time \cite{KKT,KMK}. One obtains
\begin{eqnarray}
v_b (\alpha) & \simeq & \frac{1/\sqrt{3} + \left( \alpha^2 + 2\alpha/3\right)^{1/2}}{1+\alpha}  
\label{vb}\\
u_S (\alpha) & \sim & \left( \alpha \: 
        \frac{0.72 \: \alpha+4/27 \: \sqrt{3\alpha/2}}{1+0.72 \: \alpha} \right)^{1/2}  
\label{uS}      \; ;
\end{eqnarray}
plugging these velocities into eq.~(\ref{omega_bubble}) and taking the weak transition-small
$\alpha$ limit one gets the result of ref.~\cite{KMK}, apart from the factor of order 4 discussed
in section \ref{GWs}. This turns out to be a signal too low for LISA.
Clearly a more interesting signal can be reached in the strong transition-large $\alpha$ limit:
this case could be motivated by supersymmetric extensions of the Standard Model, which in
some small regions of parameter space give rise to a very strong first order EWPT. This strong
transition, apart from creating a large amount of relic GWs, can be responsible for the
generation at the electroweak scale  of the observed baryon asymmetry 
in the Universe (see Ref.~\cite{toni} for a review).
An extensive numerical study of such EWPTs in the Minimal Supersymmetric Standard Model
and in the Next-to-MSSM has been performed in  ref.~\cite{alberto}. 
If the transition is very strong $\alpha$ is close to 1 and both velocities $v_b$ and $u_S$
reaches relativistic values;
the intensity of relic GWs from turbulence thus becomes
\be
\label{OmegaEWPT}
\alpha \simeq 1 \quad \Rightarrow \quad
h_0 ^2 \Omega_{\rm gw} (f) \simeq
5 \times 10^{-15}
\left( \frac{\rm mHz}{f} \right)^{7/2}
\left[ \frac{\beta}{H_*}  \right]^{3/2}
\left[ \frac{T_*}{100 \: \rm GeV} \right]^{7/2}
\left[ \frac{g_*}{100} \right] ^{1/4}
\; .
\ee
The stronger is the transition and the smaller is $\beta / H_*$,
because the system experiences a larger degree of supercooling. 
However, for a strong EWPT $\beta / H_* \sim 100$ is not unusual
(see ref.~\cite{TWW} for analytic estimates and ref.~\cite{alberto}
for exact numerical computations in specific examples). Such values of 
$\beta / H_*$ would give a signal above the planned sensitivity of LISA.

\subsection{MHD effects}

So far we ignored possible MHD effects which may arise because of the
presence of magnetic fields during the EWPT. This approach, however,
is not always justified since magnetic fields may have
been produced before or during the EWPT by a number of different
processes (for a review see \cite{report}). Quite solid arguments in favor
of magnetic field generation at a first order EWPT were presented by
Baym, B\"odeker and McLerran \cite{Baym}. 
Magnetic fields are seeded by the motion of 
the dipole charge layer which is produced across the bubble boundaries 
because of the small excess of top quarks over top anti-quarks and of
the potential barrier which quarks face at the bubble wall. 
Even if the seed fields are initially very tiny, they are amplified
exponentially with time by the strong turbulence which is
produced when bubble walls collide. This process is very effective due to
the high bubble wall velocities and the high value of the Reynolds number 
at that time, $\rm{Re} \sim 10^{12}$ \cite{Baym}. The amplification processes 
ends when energy equipartition is reached  between the magnetic field and the
turbulent velocity field.  We will show now that the result of Baym {\em et
al.}~may have relevant consequences for the GW signal to be expected at LISA.   

We apply here the results we derived in sec.~\ref{MHD}. We have shown there
that the MHD turbulent energy spectrum takes the same form as for the
Navier-Stokes type of turbulence upon a simple rescaling of the
parameter $\gamma$ (defined in eq.~(\ref{eps_k})) and of the wavenumber
$k_S$ (see Eqs.~(\ref{rescaling})).
Due to the huge Reynolds number we think it is a reasonable
approximation to assume that
equipartion is reached already at the maximal stirring scale,
{\em i.e.}~$k_{\rm eq} = k_S$. Furthermore, since we showed above that
$\gamma = 0$  for the EWPT with no magnetic fields, in the MHD 
we have just to replace $\gamma$ with $\delta = 1/6$. 
By replacing this value in eq.~(\ref{omega_min}) and following the same
procedure we used to derive eq.~(\ref{omega_bubble}), we get 
\be
\label{OmegaMHD}
h_0 ^2 \Omega_{\rm gw} (f) \simeq
2 \times 10^{-11}
\:
\left( \frac{\rm mHz}{f} \right)^{8/3}
u_S  ^{23/3} \: v_b^{-2/3}
\left[ \frac{\beta}{H_*}  \right]^{2/3}
\left[ \frac{T_*}{100 \: \rm GeV} \right]^{8/3}
\left[ \frac{g_*}{100} \right] ^{1/9}   \;.
\ee
This is a quite interesting result! For instance, by taking $\alpha \simeq 1$ the
velocities (\ref{vb}) and (\ref{uS}) becomes relativistic and
the overall numerical factor in eq.~(\ref{OmegaMHD}) becomes $1.8 \times 10^{-12}$:
by comparing it with 
eq.~(\ref{OmegaEWPT}) we see that, even taking into
account the smaller enhancement  due to the $2/3$ exponent for 
$ \beta/ H_*$, MHD effects amplify the signal to be expected 
at LISA by one order of magnitude. 
This signal is certainly within the planned sensitivity of the
instrument \cite{LISA}.
The strong enhancement with respect to eq.~(\ref{OmegaEWPT}) is
easily explained by the milder slope of the MHD turbulent energy
spectrum at high frequencies with respect to the Kolmogorov's.   

\begin{figure}[t]             
\begin{center}
\includegraphics[width=12cm]{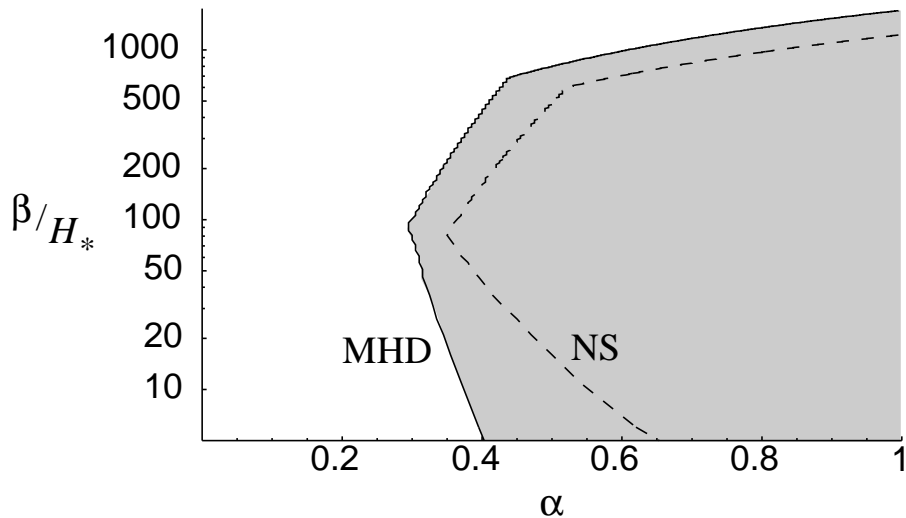}
\caption{\label{fig_LISA} The figure shows whether an EWPT characterized by a pair
$(\alpha,\beta)$ gives rise to a GW background from turbulence which 
is detectable by LISA. The shaded region is ``detectable''; the dashed and continuous lines 
refer respectively to the Navier-Stokes and to the Magneto-Hydro-Dynamical case.}
\end{center}
\end{figure}
By fixing the temperature at which EWPT happens at $100$ GeV, eqs.~(\ref{omega_bubble}) and 
(\ref{OmegaMHD}) are functions of $\alpha$, $\beta$ and $f$; the  LISA (expected) sensitivity in 
$h_0 ^2 \Omega_{\rm gw}$
clearly depends only on frequency $f$ and it is roughly $10^{-11}$ at $0.1$ mHz,
$10^{-12}$ at $1$ mHz and  $10^{-11}$ at $10$ mHz (see \cite{maggiore} for the LISA sensitivity
curve). Figure \ref{fig_LISA} shows the values of $\alpha$ and $\beta$ for which there exists
at least a range of frequency in which the signals (\ref{omega_bubble}) and (\ref{OmegaMHD})
are above the sensitivity of LISA. Clearly large values of $\alpha$, {\em i.e.}~strong transitions, are
favoured. Notice that too large values of $\beta/H_*$ are disfavoured, because in such cases the
characteristic frequency (\ref{fS_EWPT}) is too high for LISA.
Notice the effect of the MHD corrections, which substantially enhance the detectability
of the signal mostly at small values of $\beta/H_*$, because in such cases the
characteristic frequency (\ref{fS_EWPT}) is quite below the sensitivity window of LISA.

Furthermore,  we should keep in mind that tangled magnetic fields act
themselves as a source of GWs \cite{Durrer}. Assuming that magnetic
fields are in equipartition with the turbulent velocity field, this 
source should enhance the expected GW signal by a factor of two in all
the frequency range over which equipartition is established 
(see eq.~(\ref{magneticE})).

\section{Conclusions}\label{end}

In this paper we investigated the characteristics of the GW signal 
sourced by turbulence in the early universe. We extended previous work 
on the subject by considering the effects on the
turbulence energy spectrum of a continuous stirring power and of 
magnetic fields.  In the  first part of our work we derived a suitable
parameterization for the turbulent energy spectrum in the case of a 
continuous stirring power with and without external magnetic fields.
Interestingly we found that the MHD turbulent
spectrum can be written in the same form as the Navier-Stokes ones upon
a simple rescaling of only two parameters. Our treatment is original
and may find applications which go beyond the aims of this paper.
Following the approach of Kosowsky {\em et al.}~\cite{KMK}, who consider only 
Kolmogorov type of turbulence, we used our formulation to determine the 
characteristics of the GW signal produced by cosmological turbulence
with a more general energy spectrum.
In the Kolmogorov case we found minor discrepancies with the result
reported in \cite{KMK}.  
Deviations from the Kolmogorov spectrum may have relevant
consequences for the GW signal. Indeed, since the modified spectra are
generally less steep than the Kolmogorov's, these deviations will result in a
stronger signal at high frequencies. Furthermore, our analysis may
allow to extract valuable informations about the nature of the 
turbulence source and the presence of primordial magnetic fields  
from the GW background power spectrum which may be measured by forthcoming
experiments. 

We applied our results to estimate the GW expected signal for two 
possible kind of sources. 
In sec.~\ref{neutrino} we considered GW production by neutrino 
inhomogeneous diffusion according to the
mechanism proposed in Ref.~\cite{dario}. We showed that a detectable
signal can be produced only if the amplitude of the lepton number
fluctuations is close to unity over a wide range of wavenumbers.
This possibility is not unreasonable as active-sterile neutrino
oscillations or some models of leptogenesis based on the 
Affleck-Dine scenario can indeed give rise to domains
with opposite lepton number.  
Although we argued that the former scenarios 
can hardly give rise to GW signal detectable by LISA, the latter
scenario offers more promising observational perspectives. 

Finally in sec.~\ref{EWPT} we considered GW production 
by turbulence at the end of a first
order phase transition. In the absence of strong magnetic fields no
substantial deviations have to be expected from the results of 
Ref.~\cite{KMK} based on the assumption of a Kolmogorov turbulence
spectrum. In the case of the EWPT  the GW signal
may be above the LISA planned sensitivity only if the transitions
is very strong with large bubble wall velocities. This may be achieved
in next to minimal extensions of the supersymmetric standard model
\cite{alberto}. A more favorable situations may be obtained 
if strong magnetic fields were present at the end of the transition.
This is actually to be expected in the case of the EWPT if it is first
order \cite{Baym}. We showed that in this case the signal can be 
strongly enhanced with respect to the non-magnetic case and be detectable
by LISA if bubble wall velocity was not too much smaller than unity.
Similar consideration may apply to other physical situations like, 
for example, during reheating at the end of inflation.
These results open new perspectives for a successful detection of 
GWs backgrounds produced in the early universe.

\subsection*{Acknowledgments} 

The authors thanks G. Falkovich for useful comments.
D.~G.~thanks the CERN theory division for ospitality during
the last writing of this paper.

\appendix
\section*{Appendix}
\section*{Computation of the spectrum $ h_0 ^2 \Omega_{\rm gw} (f)$}

In this appendix we work out the detailed computation of the relic spectrum
$h_0 ^2 \Omega_{\rm gw} (f)$ in the three cases $\gamma < 2/3$, $\gamma \simeq 2/3$ and $\gamma>2/3$,
along the strategy described in section \ref{GWs}.

If $\gamma<2/3$ eq.~(\ref{k_41}) becomes
\be
\langle \: h_{ij} ({\bf k}, t_*) \: h_{ij} ^* ({\bf k} ', t_*)
\: \rangle \simeq
\frac{27 \cdot 64\sqrt{2}\: \pi^7 (G w)^2}{(2 -3 \gamma) V^2}\: \Delta t_* \:
   \: u_S^4 \: k_S ^{2/3-\gamma} \:
\delta^3 ({\bf k} - {\bf k} ')  \: k^{ -20/3 + \gamma}   \; ,
\ee
which in real space reads
\begin{eqnarray} 
\langle \: h_{ij} ({\bf x}, t_*) \: h_{ij} ({\bf x}, t_*)
\: \rangle & \simeq &
\frac{27 \cdot 4\sqrt{2}\:\pi^2\: (G w)^2}{2-3\gamma}\: \Delta t_* \:
  \: u_S^4 \: k_S ^{2/3-\gamma}
\int^{k_D} _{k_S}  k^{-14/3+\gamma} d k         \\
& = &
\frac{27 \cdot 4\sqrt{2}\:\pi^2\: (G w)^2}{2-3\gamma}\: \Delta t_* \:
  \: u_S^4 \: k_S ^{-3}
\int^{k_D} _{k_S}  \left( \frac{k}{k_S} \right)^{-\frac{11-3\gamma}{3}} \frac{d k}{k}
\label{real}\; .
\end{eqnarray}
Changing variable from the turbulent scale $k$ to the characteristic frequency $f$
(see eq.~(\ref{freq})), we have
\be
\langle \: h_{ij} ({\bf x}, t_*) \: h_{ij} ({\bf x}, t_*)
\: \rangle  \simeq
\frac{81 \cdot 8\sqrt{2}\:\pi^2 \:  (G w)^2}{(2-3\gamma)(4+3\gamma)}
\: \Delta t_* \:  
\: u_S ^4
\: k_S ^{-3}
\int^{f_D} _{f_S}  \left( \frac{f}{f_S} \right)^{-\frac{2(11-3\gamma)}{4+3\gamma}} \frac{df}{f}      \; ,
\label{real_mag}
\ee
where $f_S \equiv f_{k_S}$ is the characteristic frequency of the largest length scale.
The characteristic amplitude squared is thus
\be
h_c ^2 (f, t_*) =  
\frac{81 \cdot 4\sqrt{2}\:\pi^2 \:  (Gw)^2}{(2-3\gamma)(4+3\gamma)}
\: \Delta t_*
\: u_S ^4
\: k_S ^{-3}
\left( \frac{f}{f_S} \right)^{-\frac{2(11-3\gamma)}{4+3\gamma}}
\; .
\ee

The characteristic amplitude squared measured today at a frequency $f$ (see eq.~(\ref{red})) is
\be
h_c ^2 (f,t_0)  \simeq   \frac{3.3 \times 10^{-18}}{(2-3\gamma)(4+3\gamma)} 
\: u_S ^4
\left[ \frac{\Delta t_*}{H_* ^{-1}} \right]
\left[ \frac{k_S}{H_*} \right]^{-3}
\left[ \frac{T_*}{\rm MeV} \right]^{-2}
\left[ \frac{g_*}{10.75} \right] ^{-2/3}
\left( \frac{f}{f_S} \right)^{-\frac{2(11-3\gamma)}{4+3\gamma}}
\label{hc} \; ,
\ee
where
we used $G w   = H_*/(2 \pi)$ and both frequencies $f$ and $f_S$ have been redshifted.
By means of eqs.~(\ref{relaz}) and (\ref{fS}) we finally have the spectrum, 
eq.~(\ref{omega_min}).

The above computations can be easily generalized to the $\gamma>2/3$ case:
\be
\langle \: h_{ij} ({\bf k}, t_*) \: h_{ij} ^* ({\bf k} ', t_*)
\: \rangle \simeq
\frac{27 \cdot 64\sqrt{2}\: \pi^7 (G w)^2\: (11-3\gamma)}{(3\gamma-2)(13-6\gamma) V^2}
\: \Delta t_* \:
  \:
u_S ^4 \: k_S ^{4/3 - 2\gamma}  \:
\delta^3 ({\bf k} - {\bf k} ')  \: k^{-\frac{2(11-3\gamma)}{3}}  \; ,
\ee
{\em i.e.}
\begin{eqnarray}
\langle \: h_{ij} ({\bf x}, t_*) \: h_{ij} ({\bf x}, t_*)
\: \rangle
\! & \! \simeq \! & \!
\frac{27 \cdot 4\sqrt{2}\: \pi^2 (G w)^2\: (11-3\gamma)}{(3\gamma-2)(13-6\gamma}
\: \Delta t_* \:
  \:
u_S ^4 \: k_S ^{-3}
\int^{k_D} _{k_S}  \left( \frac{k}{k_S} \right)^{-\frac{13-6\gamma)}{3}} \frac{d k}{k}     \nonumber \\
\! & \! \simeq \! & \!
\frac{81 \cdot 8\sqrt{2}\: \pi^2 (G w)^2\: (11-3\gamma)}{(3\gamma-2)(13-6\gamma)(4+3\gamma)}
\: \Delta t_* \:
 \:  u_S ^4
\: k_S ^{-3}
\int^{f_D} _{f_S}   \left( \frac{f}{f_S} \right)^{-\frac{2(13-6\gamma)}{4+3\gamma}} 
\frac{d f}{f}  \label{real_min} \\
h^2 _c (f, t_*) \! & \! \simeq \! & \!
\frac{81 \cdot 4\sqrt{2}\: \pi^2 (G w)^2\: (11-3\gamma)}{(3\gamma-2)(13-6\gamma)(4+3\gamma)}
\: \Delta t_* \:
 \: u_S ^4
\: k_S ^{-3}
\: \left( \frac{f}{f_S} \right)^{-\frac{2(13-6\gamma)}{4+3\gamma}}
\; .
\end{eqnarray}

The characteristic amplitude measured today is
\be
h_c ^2 (f) \simeq   \frac{3.3 \times 10^{-18} \: (11-3\gamma)}{(3-\gamma)(13-6\gamma)(4+3\gamma)} 
\: u_S ^4
\left[ \frac{\Delta t_*}{H_* ^{-1}} \right]
\left[ \frac{k_S}{H_*} \right]^{-3}
\left[ \frac{T_*}{\rm MeV} \right]^{-2}
\left[ \frac{g_*}{10.75} \right] ^{-2/3}
\left( \frac{f}{f_S} \right)^{-\frac{2(13-6\gamma)}{4+3\gamma}}
\; ,
\ee
which, combined with eq.~(\ref{relaz}), gives eq.~(\ref{omega_mag}).

Looking at eq.~(\ref{casi}) we see that the intermediate case $\gamma \simeq 2/3$
is formally the same as $\gamma < 2/3$ case,
with the replacements $\gamma \to 2/3$ and $\frac{3}{2-3 \gamma} \to 1$,
and with an extra logarithmic factor.
The present-day observable are thus
\be
h_c ^2 (f)  \simeq  1.8 \times 10^{-19} 
\: u_S ^4
\left[ \frac{\Delta t_*}{H_* ^{-1}} \right]
\left[ \frac{k_S}{H_*} \right]^{-3}
\left[ \frac{T_*}{\rm MeV} \right]^{-2}
\left[ \frac{g_*}{10.75} \right] ^{-2/3}
\left( \frac{f}{f_S} \right)^{-3}
 \log{\frac{f}{f_S}}
\ee
and the spectrum given in eq.~(\ref{omega_ug}).

As a cross-check, notice that the $\gamma \simeq 2/3$ case is also equal to
the $\gamma<2/3$ one, again modulo proper replacements and the
logarithmic factor. The present-day observable computed in the two ways are equal.


\end{document}